\documentclass[prd,preprint]{revtex4}
\usepackage{amsmath,amssymb,latexsym}

\newcommand{\bea}{\begin{eqnarray}}
\newcommand{\eea}{\end{eqnarray}}
\newcommand{\ba}{\begin{array}}
\newcommand{\ea}{\end{array}}
\newcommand{\be}{\begin{equation}}
\newcommand{\ee}{\end{equation}}
\newcommand{\beas}{\begin{eqnarray*}}
\newcommand{\eeas}{\end{eqnarray*}}

\newcommand{\nbox}{{\,\lower0.9pt\vbox{\hrule \hbox{\vrule height 0.2
cm \hskip
0.2 cm \vrule height 0.2 cm}\hrule}\,}}

\DeclareFixedFont{\xiiss}{OT1}{cmss}{m}{n}{12}
\DeclareFixedFont{\ixss}{OT1}{cmss}{m}{n}{9}
\DeclareFixedFont{\cmrnine}{OT1}{cmr}{m}{n}{9}

\newcommand{\CC}{\hbox{\xiiss C\kern-.4emI}}
\newcommand{\RR}{\hbox{\xiiss R\kern-.45emI}}
\newcommand{\ZZ}{\hbox{\xiiss Z\kern-.4emZ}}
\newcommand{\CCs}{\hbox{\ixss C\kern-.4emI}}
\newcommand{\ZZs}{\hbox{\ixss Z\kern-.4emZ}}
\newcommand{\pa}{\partial}

\newcommand{\pasl}{\pa\kern-.55em /}
\def\href#1#2{#2}

\begin{document}
\title{\Large \bf De Sitter Invariant Vacuum States, Vertex Operators,
and Conformal Field Theory Correlators  }
\author{Aharon Casher$^{a,b}$, Pawel O. Mazur$^a$ and Andrzej J. Staruszkiewicz$^c$}
\address{$^a$ Department of Physics and Astronomy \\
University of South Carolina \\
\sl Columbia, S.C. 29208, U.S.A. \\
{\tt mazur@mail.psc.sc.edu} }
\address{$^b$ School of Physics and Astronomy \\ Tel Aviv University \\ Tel Aviv 69978, Israel
\\
{\tt ronyc@post.tau.ac.il} }
\address{$^c$ Marian Smoluchowski Institute of Physics \\
Jagellonian University \\
Reymonta 4, 30-059 Krak\'{o}w, Poland \\
{\tt astar@th.if.uj.edu.pl} }

\abstract{We show that there is only one physically acceptable
vacuum state for quantum fields in de Sitter space-time which is
left invariant under the action of the de Sitter-Lorentz group
$SO(1,d)$ and supply its physical interpretation in terms of the
Poincare invariant quantum field theory (QFT) on one dimension
higher Minkowski space-time. We compute correlation functions of
the generalized vertex operator $:e^{i\hat{S}(x)}:$, where
$\hat{S}(x)$ is a massless scalar field, on the $d$-dimensional de
Sitter space and demonstrate that their limiting values at
time-like infinities on de Sitter space reproduce correlation
functions in $(d-1)$-dimensional Euclidean conformal field theory
(CFT) on $S^{d-1}$ for scalar operators with arbitrary real
conformal dimensions. We also compute correlation functions for a
vertex operator $e^{i\hat{S}(u)}$ on the \L obaczewski space and
find that they also reproduce correlation functions of the same
CFT. The massless field $\hat{S}(u)$ is the nonlocal transform of
the massless field $\hat{S}(x)$ on de Sitter space introduced by
one of us. }}

\pacs{04.62.+v, 11.10.Kk, 11.25.Hf, 11.30.Cp, 11.90.+t \qquad
\qquad }
\date{\today}
\maketitle

\section{Introduction.} The maximally symmetric and homogeneous
solution to Einstein's equations with the positive vacuum energy
density is the de Sitter space-time. It is therefore important to
understand the behavior of quantum fields on this space. Recently
this subject has been taken up with renewed interest after the new
and spectacular evidence for the positive cosmological constant
and the accelerating expansion of the Universe has turned up in
recent cosmological measurements. Although quantum field theory
(QFT) on de Sitter space-time is a rather well studied subject
\cite{as87,as89,as90,as92,as77,as02,as04,cherntag,tag,mott,wyroz,lokas,amm97,mazmott,roman87,strom}
there are some aspects of it which attracted considerable interest
most recently \cite{amm97,mazmott,strom}. One of them is the
question of uniqueness of the de Sitter invariant vacuum state
\cite{as87,as89,as90,as92,as02,as04,cherntag,tag,mott,wyroz,lokas,strom}
while the second one is the fact first established in
\cite{as87,as89,tag,amm97,mazmott} that the limiting values of
correlation functions at time-like infinities on de Sitter space
reproduce correlation functions of Euclidean CFT in one dimension
lower. This property is now referred to as the de Sitter/CFT
correspondence in the recent literature \cite{strom}. In some
early papers on the subject of QFT on de Sitter space an analysis
of solutions to the wave equation for a massive scalar field led
to the conclusion that there exists a one-parameter family of de
Sitter invariant vacuum states for a massive scalar field
$\hat{\Phi}(x)$ \cite{cherntag,tag,mott}. In the current
literature the origin of this one-parameter family of vacuum
states is regarded as mysterious \cite{cherntag,tag,mott,strom}.
Similarly, the proposal of de Sitter/CFT correspondence based on
the analysis of the propagator for massive fields on de Sitter
space leads to some difficulties because the scaling dimensions

\be \Delta_{\pm} = \frac{d - 1}{2} {\pm} \sqrt{(\frac{d-1 }{2})^2
- m^2} \ee

\noindent become complex for $m^2 > (\frac{d-1}{2})^2$
\cite{tag,amm97,strom}.

\vskip .1cm \noindent The purpose of this paper is to address
these two aspects of QFT on de Sitter space. In particular, we
shall demonstrate the existence of the one-parameter family of de
Sitter invariant vacuum states $\mid \lambda>$ by solving
equations following from the conditions that they are annihilated
by generators of the de Sitter group, $\hat{M}_{_{ab}}\mid
\lambda> = 0$ and we will connect their origin to the same
property of the Lorentz covariant QFT on Minkowski space. It is
well known that the spatial infinity of $d+1$-dimensional
Minkowski space-time is the $d$-dimensional de Sitter space
\cite{pgb}. Quantum field theory of massless and massive fields on
de Sitter space inherits its vacuum structure from the Lorentz
invariant QFT of massless fields in Minkowski space-time. There
exists a one-parameter family of Lorentz invariant vacuum states
for QFT on Minkowski space-time. This implies the presence of
$\lambda$-vacua for QFT on de Sitter space. It turns out that
these states are physically unacceptable for $\lambda \neq 0$ as
they introduce correlations between the antipodal points on de
Sitter space-time. We shall show that there exists only one
physically acceptable and unique de Sitter invariant vacuum state
for massive and massless fields on de Sitter space, corresponding
to $\lambda = 0$, and supply the physical interpretation of this
state in terms of the Poincare invariant QFT of a massless
particle in $d+1$-dimensional Minkowski space-time
\cite{as87,as89,as90,as92,as02,as04,wyroz,lokas}. The positive
frequency modes $e^{-ik\cdot x }$ of a massless particle in
Minkowski space-time properly projected on de Sitter space lead to
the unique positive frequency modes on de Sitter space
\cite{as87,as89,as90,as92,as02,as04,wyroz,lokas} and to a unique
physically acceptable de Sitter invariant vacuum state.

We shall also address the now well known fact
\cite{tag,as87,as89,amm97,mazmott,strom} that behavior of
two-point correlation functions at time-like infinities on $d$
dimensional de Sitter space (and at spatial infinity on \L
obaczewski space) is characteristic of the $d-1$ dimensional
Euclidean conformal field theory (CFT). In order to make the de
Sitter/CFT correspondence as transparent as possible and at the
same time satisfy the requirement that conformal dimensions
$\Delta$ of scalar operators in CFT be real and positive we shall
introduce a family of generalized vertex operators
$e^{i\hat{S}(x)}$ for a massless scalar field $\hat{S}(x)$ and
compute their two-point correlation functions
\cite{as87,as89,as90,as92,as02,as04}. We demonstrate that every
correlation function for a scalar operator of conformal dimension
$\Delta$ is reproduced by our vertex operators on de Sitter
space-time.

\section{De Sitter invariant vacuum states.}

De Sitter space is a single-sheeted hyperboloid in the $d+1$
dimensional Minkowski space:

\be x^a\eta_{ab}x^b = {x {\cdot} x} = -1  \ . \ee

\noindent This representation of de Sitter space makes it
transparent that the de Sitter isometry group is $SO(1,d)$. The
generators of this symmetry corresponding to Killing vectors on de
Sitter space are

\be M_{ab} = i(x_a\partial_b - x_b\partial_a)  \ . \ee

\noindent Using the standard global parametrization of the
hyperboloid

\be x^0 = sinh\tau \ \ \ , \ \ \  x^i = n^i cosh\tau \ \ \ ,  \ \
\ \textbf{n}^2 = 1  \ , \ee

\noindent one obtains the induced metric on de Sitter space:

\be ds^2 = g_{\mu\nu}dx^{\mu}dx^{\nu} = d\tau^2 - cosh^2\tau
d\Omega^2_{d-1} \ . \ee

\noindent The d'Alambertian on de Sitter space is

\be {{\Box}_g} =
|det(g)|^{-\frac{1}{2}}\partial_{\mu}(|det(g)|^{\frac{1}{2}}g^{\mu\nu}\partial_{\nu})
= (cosh\tau)^{1-d}\partial_{\tau}((cosh\tau)^{d-1}\partial_{\tau})
+ cosh^{-2}\tau \textbf{L}^2  \ , \ee

\noindent where $\textbf{L}^2$ is the Laplacian on the unit
$S^{d-1}$ sphere, and it is the same as the Casimir operator
${\cal{C}}=\frac{1}{2}M_{ab}M^{ab}$ of the de Sitter group. The
Casimir operator ${\cal{C}}$ enters a very useful identity
relating the d'Alambertian on Minkowski space ${\Box}_{d+1} =
\eta^{ab}\partial_{a}\partial_{b}$ to the one on de Sitter space

\be \label{identity} {\cal{C}} = \frac{1}{2}M_{ab}M^{ab} = - x^2
{\Box}_{d+1} + {\cal E}({\cal E} + d - 1)  \ , \ee

\noindent where ${\cal E} = x^a\partial_a$ is the Euler operator
(generator of dilatation subgroup of the conformal group). This
identity allows us to describe massive (and massless) scalar
fields on de Sitter space in terms of a massless scalar field on
Minkowski space. A massless scalar field $\Phi_{_\Delta}(x)$ in
Minkowski space with the scaling dimension $\Delta$

\be \Phi_{_\Delta}(\lambda x) = \lambda^{-\Delta}
\Phi_{_\Delta}(x)  \ \ \  ,  \ \ \  {\cal{E}}\Phi_{_\Delta}(x) = -
\Delta\Phi_{_\Delta}(x)  \ , \ee

\noindent corresponds to a massive scalar field on de Sitter space
with a mass $m$ such that

\be m^2 = \Delta(d - 1 - \Delta)  \ . \ee

\noindent For a given mass there are two scaling dimensions

\be \Delta_{\pm} = \frac{d - 1}{2} {\pm} \sqrt{(\frac{d-1 }{2})^2
- m^2} \ . \ee

\noindent In the following we choose $\Delta = \Delta_{-}$ for the
scaling dimension of a massless scalar field on Minkowski space
corresponding to a massive scalar on de Sitter space. The dynamics
of this massive scalar field $\Phi(x)$ is defined by the following
Lagrangian density:

\be {\cal {L}} =
\frac{1}{2}(g^{\mu\nu}\partial_{\mu}\Phi\partial_{\nu}\Phi -
m^2\Phi^2)  \ . \ee

\noindent The Poincare invariant QFT of a massless scalar field on
Minkowski space induces a de Sitter invariant QFT of a massive (or
massless) scalar field on de Sitter space with a unique vacuum
state. This follows immediately from (\ref{identity}) and the
observation that the sign of the Klein-Gordon norm for a massless
scalar field on Minkowski space is preserved upon projection on
the de Sitter hyperboloid \cite{as87,as89,as90,as92,as02,as04}.
This has two immediately clear implications. First, a positive
frequency solution for a massless scalar on Minkowski space

\be \label{posfreq} \Phi^{(+)}_{_\Delta}(x) = \int
\frac{(dk)_{d}}{k_0} e^{-i{k{\cdot}x}} a_{_\Delta}(k) \ , \ee

\noindent where

\be a_{_\Delta}(\lambda k) =
\lambda^{-\tilde{\Delta}}a_{_\Delta}(k) \ \ \  ,  \ \ \
\tilde{\Delta} = d -1 - \Delta \ , \ee

\noindent and $k$ is a null momentum vector $k^2 = 0$, when
projected on the de Sitter hyperboloid ${x{\cdot}x} = -1$ becomes
a positive frequency solution for the Klein-Gordon wave equation
on de Sitter space. Choosing

\be a_{_\Delta}(k) = {\sum_{nM}} N_n
{k_0}^{-\tilde{\Delta}}Y_{nM}(\hat{k}) a_{nM} \ , \ee

\noindent with $\tilde{\Delta} = \Delta_{+}$, $\hat{k} =
\frac{k}{k_0}$, and $Y_{_{nM}}(\hat{k})$ the spherical harmonics
on $S^{d-1}$, one obtains the positive frequency modes of the
massive scalar field on de Sitter space

\be \label{mode} f_n(\tau)Y_{nM}(\textbf{n}) = N_n \int
\frac{(dk)_{d}}{k_0} e^{-i{k{\cdot}x}}
{k_0}^{-\tilde{\Delta}}Y_{nM}(\hat{k})  \ . \ee

\noindent The integral (\ref{mode}) can be easily evaluated
\cite{lokas} and it is given below in (\ref{modes}) with

\be \label{coeffs} A^{(0)}_n =
\frac{1}{2}\sqrt{\frac{\Gamma(\frac{n +
{\Delta}_{+}}{2})\Gamma(\frac{n +
{\Delta}_{-}}{2})}{\Gamma(\frac{n + 1 +
{\Delta}_{+}}{2})\Gamma(\frac{n + 1 + {\Delta}_{-}}{2})}} \,\,\,\
, \,\,\ B^{(0)}_n = - \frac{i}{2A^{(0)}_n}  \,\ . \ee

\noindent The spherical harmonics $Y_{nM}(\hat{k})$ and
$Y_{nM}(\textbf{n})$ are eigenfunctions of the Laplacian on
$S^{d-1}$ corresponding to the eigenvalue $n(n+d-2)$ and the
degeneracy

\be d_n(d) = \frac{(2n + d-2)}{(d-2)!}(n + 1)...(n + d -3)  \ .
\ee

\noindent Second, by promoting the amplitudes $a_{nM}$ to the
annihilation operators $\hat{a}_{nM}$ one also obtains the
positive frequency part of the massive scalar field operator
$\hat{\Phi}^{(+)}(x)$ on de Sitter space. The unique de Sitter
invariant vacuum state $\mid 0>$ is defined by the condition

\be \hat{\Phi}^{(+)}(x)\mid 0> = 0  \ .  \ee

\noindent We now construct the generators ${\hat{M}_{ab}}$ of the
de Sitter group acting on the Fock space of a massive scalar field
$\hat{\Phi}(x)$. To claim de Sitter invariance of a vacuum state
one needs to impose the conditions

\be {\hat{M}_{ab}}\mid 0> = 0 \ , \ee

\noindent where conserved objects

\be {\hat{M}_{ab}} = \int (cosh\tau)^{d-1} d\Omega_{d-1}
:\hat{T}^0_{\mu}(x):\xi^{\mu}_{ab} \ , \ee

\noindent are the generators of the de Sitter group evaluated at
$\tau = 0$, $\xi^{\mu}_{ab}$ are the de Sitter space Killing
vectors, and

\be :\hat{T}_{\mu\nu}(x): =
:\partial_\mu\hat{\Phi}(x)\partial_{\nu}\hat{\Phi}(x): -
\frac{1}{2} g_{\mu\nu}:{\cal{L}}(x): \ , \ee

\noindent is the stress-energy-momentum tensor of a massive scalar
field on de Sitter space. The field operator of a massive scalar
field on the de Sitter space is

\be \hat{\Phi}(x) = {\sum_{nM}} (f_n(y)Y_{nM}(\textbf{n})
\hat{a}_{nM} + h.c.)  \ , \ee

\noindent where

\bea f_n(y) &=& (1 - y)^{s_{-}}[A_n F(s_{-}+ \frac{n}{2},s_{-} +
\frac{2 -d -n}{2};\frac{1}{2};y) + \,\nonumber\\
&& + B_n y^{\frac{1}{2}}F(s_{-}+ \frac{n +1}{2},s_{-}+ \frac{3 - d
- n}{2};\frac{3}{2};y)] \,, \label{modes}\eea

\noindent $s_{-} = \frac{1}{2}\Delta_{-}$, $y = tanh^2\tau$, and
$F$ is the hypergeometric function. The coefficients $A_n$ and
$B_n$ satisfy the condition

\be i(\overline{A}_nB_n - A_n\overline{B}_n) = 1  \ , \ee

\noindent following from the positivity of the Klein-Gordon norm
for the positive frequency modes $f_n(y)Y_{nM}(\textbf{n})$. To
find the de Sitter invariant vacuum state it is sufficient to
compute the de Sitter group generators ${\hat{M}_{0i}}$ and demand
that they annihilate this vacuum state. The Killing vectors
$\xi^{\mu}_{ab}$, $a=0$, $b=i$, corresponding to these generators
are expressed in terms of the $n =1$ spherical harmonics,

\be \xi_{0} = Y_{1M_0}(\textbf{n})  \ \ \  ,  \ \ \xi_{\alpha} = -
sinh{\tau}cosh{\tau}
\partial_{\alpha}Y_{1M_0}(\textbf{n})  \ . \ee

\noindent We obtain the following formula for the only
non-vanishing terms in the normal ordered expression for the
generators ${\hat{M}_{0i}}$ acting on the vacuum state $\mid 0>$

\be {\frac{1}{2}}{\sum_{nM}} {\sum_{n'M'}} C_{nn'} {\int
d\Omega_{d-1}} Y_{1M_0} \overline{Y}_{nM}
\overline{Y}_{n'M'}{\hat{a}^{\dagger}}_{nM}{\hat{a}^{\dagger}}_{n'M'}
+ ...
 \ , \label{generators}\ee

\bea C_{nn'} &=&
\overline{B}_{n}\overline{B}_{n'} + (m^2+{\frac{1}{2}}n(n+d-2)+ \,\nonumber\\
&& + {\frac{1}{2}}n'(n'+d-2)-{\frac{1}{2}}(d-1))\overline{A}_{n}
\overline{A}_{n'} \,. \label{coefs}\eea

\noindent The generators $\hat{M}_{0i}$ annihilate the vacuum
state $\mid 0>$ only when the coefficients in front of two
creation operators in (\ref{generators}) vanish. The
Clebsch-Gordan coupling coefficients for $SO(d)$ do not vanish for
$\mid n - n'\mid = 1$. This implies that $C_{nn'}$ for $\mid n -
n'\mid = 1$ must vanish. We obtain the following simple equation
for $\alpha_n = i\frac{B_n}{A_n}$

\be \label{secondcontraint} \alpha_n \alpha_{n+1} = m^2 + n(n+d-1)
\ , \ee

\noindent which together with the K-G norm condition leads to the
following result: for $n$ even,

\be A_n = e^{\lambda}A^{(0)}_n  \ \ \  ,   \ \ \ B_n =
e^{-\lambda}B^{(0)}_n  \ , \ee

\noindent and for $n$ odd,

\be A_n = e^{-\lambda}A^{(0)}_n  \ \ \  ,  \ \ \ B_n =
e^{\lambda}B^{(0)}_n  \ , \ee

\noindent with $\lambda$ real. $A^{(0)}_n$ and $B^{(0)}_n$ are the
coefficients obtained before (\ref{coeffs}) from the evaluation of
the integral (\ref{mode}). The closure of the Lorentz-de Sitter
Lie algebra implies that the remaining generators will also
annihilate the vacuum state. There exists a simple canonical
transformation between the field operator
$\hat{\Phi}_{\lambda}(x)$ corresponding to the modes (\ref{modes})
and the field operator $\hat{\Phi}(x)$ corresponding to the unique
vacuum state (\ref{mode}):

\be \hat{\Phi}_{\lambda}(x) = cosh{\lambda}\hat{\Phi}(x) +
sinh{\lambda}\hat{\Phi}(-x) \ , \ee

\noindent which makes the correlations between antipodal points
transparent. The modes (\ref{modes}) correspond to the following
Lorentz, but not Poincare, invariant choice for positive frequency
modes for a QFT on Minkowski spacetime:

\be f_{k}(x) = cosh{\lambda}e^{-ik{\cdot}x} +
sinh{\lambda}e^{ik{\cdot}x} \ . \ee

\noindent This concludes our demonstration of the existence of a
one-parameter family of de Sitter invariant vacuum states $\mid
\lambda>$ and elucidates their physical meaning.

\section{Vertex operators on de Sitter space-time \\ and their
correlation functions.}

The special case of a massless scalar field which is the phase
$\hat{S}(x)$ canonically conjugate to the electric charge operator
$\hat{Q}$ has been studied extensively in the physically most
interesting case $d=3$ in \cite{as87,as89,as90,as92,as02,as04}. It
turns out that a special care is needed in the treatment of the
zero mode $n=0$ of the Laplacian on the unit sphere. Indeed, our
formula for $A^{(0)}_n$ has a simple pole at $n=0$ for a massless
field and as such is meaningless. We need to solve the d'Alambert
equation for the $n=0$ part of a massless scalar field which from
now on we denote $\hat{S}(x)$. The dynamics of $\hat{S}(x)$ is
derived from the somewhat more conveniently normalized Lagrangian
density

\be {\cal{L}} =
(2e^2{\Omega}_{d-1})^{-1}g^{\mu\nu}\partial_{\mu}\hat{S}\partial_{\nu}\hat{S}
 \ , \ee

\noindent where ${\Omega}_{d-1}$ is the volume of $S^{d-1}$ and
$e$ is the coupling constant whose meaning is that of a unit of
electric charge in the $d = 3$ case. We find the following
expression for the massless field operator $\hat{S}(x)$

\be \label{massless} \hat{S}(x) = \hat{S}_0 + e\hat{Q}f_0(\tau) +
{\sum_{nM}}(f_n(\tau)Y_{nM}(\textbf{n})\hat{a}_{nM} + h.c.) \, \ee

\noindent where the sum is over $n\neq 0$ and the zero mode
$f_0(\tau)$ satisfies the following equation:

\be (cosh{\tau})^{d-1}\partial_{\tau}f_0(\tau) = 1 \ . \ee

\noindent The operators $\hat{S}_0$ and $\hat{Q}$ are canonically
conjugate:

\be [\hat{S}_0,\hat{Q}] = ie \ . \ee

\noindent The annihilation and creation operators satisfy the
following commutation relations:

\be [\hat{a}_{nM},\hat{a}^{\dag}_{n'M'}] =
e^2\Omega_{d-1}\delta_{nn'}\delta_{MM'}  \ . \ee

\noindent It is easy to show that the de Sitter group generators
evaluated for a massless scalar $\hat{S}(x)$ do not depend on the
constant phase $\hat{S}_0$ but do depend on the charge operator
$\hat{Q}$. The de Sitter invariant vacuum state is then
characterized by the following conditions:

\be \hat{Q}\mid{0}> = 0  \ \ \  ,  \ \  \hat{a}_{nM}\mid{0}> = 0 \
. \ee

\noindent In the representation in which the operator
$e^{i\hat{S}_0}$ is diagonal and eigenstates of $\hat{Q}$ are
normalizable $S_0$ is a periodic variable with the usual period of
$2{\pi}$. It is a phase conjugate to a charge $\hat{Q}$.

It is convenient, following \cite{as87,as89,as90,as92,as02,as04},
to introduce the generalized vertex operator on de Sitter space

\be \hat{V}(x) = :e^{i\hat{S}(x)}: \ , \ee

\noindent and compute its two-point correlation function

\be \label{correlator} <0\mid
:e^{i\hat{S}(x)}::e^{-i\hat{S}(y)}:\mid 0> = e^{F(x,y)} \ , \ee

\noindent where

\be F(x,y) = e^2(<0\mid \hat{S}^{+}(x)\hat{S}^{-}(y)\mid 0> -
\frac{i e^2}{2}(f_0(x) - f_0(y)))  \ , \ee

\noindent with $\hat{S}^{+}(x)$ and $\hat{S}^{-}(x)$ the positive
and negative frequency parts of $\hat{S}(x)$. The master function
$F(x,y)$ can be computed exactly but its exact form is of no
concern for us here. The remarkable property of this function is
its universal asymptotic behavior when $\tau(x) = \tau(y) =
{\pm}\infty$. One finds in this limit

\be \label{conf} F(\hat{x},\hat{y}) = - C(d)e^2 \left(
\frac{1}{(d-3)!} ln\Bigl(\frac{1-cos\theta}{2}\Bigr) +
C'(d)\right) \ , \ee

\noindent where

\be cos\theta = \textbf{n}(x)\cdot\textbf{n}(y) \ , \ee

\noindent and

\be C(d) = {{2^{d-3}\Gamma^2({{d-1}\over 2})}\over {(d-2)\pi}}  \
, \ee

\be C'(d) = (1 + (-1)^d)ln2 + (-1)^d{\sum_{n=1}^{d-2}}(-1)^n
\frac{1}{n} \ . \ee

\noindent This implies that the two-point correlation function of
the vertex operator on time-like infinity $S^{d-1}$ is equal to
the two-point correlation function of the scalar operator with the
conformal dimension $\Delta = C(d)e^2$ in the Euclidean CFT on
$S^{d-1}$. The underlying reason for this relation is the
isomorphism of the conformal group of $R^{d-1}$ (or $S^{d-1}$) and
the de Sitter group $SO(1,d)$ on $d$-dimensional de Sitter space
which is the basic kinematical fact with many different dynamical
realizations \cite{as87,as89,pom90,pom91,amm96,amm97,mazmott}. Our
basic formula (\ref{correlator}) generalizes the one obtained in
the special case of $d=3$ \cite{as87,as89}. Unlike the case of the
two-point correlation function of a massive scalar field first
discussed in \cite{tag,amm97} there are no problems with the
vanishing with $\tau\rightarrow\infty$ of an overall amplitude of
the $d-1$-dimensional CFT two-point function and the complex
values of the scaling dimensions $\Delta_{{\pm}}$ for $m^2
> (\frac{d-1}{2})^2$. Using our vertex operators we can reproduce
all multi-point correlation functions of scalar primary scaling
operators of any scaling dimension in the Euclidean
$d-1$-dimensional CFT.

\section{Correlation functions on the \L
obaczewski space.}

The operator $\hat{S}_0$ can be extracted from $\hat{S}(x)$ by
averaging over the sphere $S^{d-1}$, $\tau = 0$, which is an
intersection of $x^0 = 0$ with the de Sitter hyperboloid
$x{\cdot}x = -1$. The proper Lorentz invariant generalization of
$\hat{S}_0$ is the following nonlocal transform from de Sitter
space to the \L obaczewski space of velocities $u$
\cite{as87,as89,as90,as92,as02,as04}

\be \label{StarTransform} \hat{S}(u) = \frac{2}{\Omega_{d-1}} \int
(dx)_{d+1}\delta(x{\cdot}x + 1) \delta(u{\cdot}x) \hat{S}(x) \ ,
\ee

\noindent where $u{\cdot}u = 1$. The induced metric on the \L
obaczewski space is

\be ds^2 = d{\psi}^2 + sinh^2{\psi}d\Omega^2_{d-1} \ . \ee

\noindent One can show that $\hat{S}(u)$ satisfies the Laplace
equation on \L obaczewski space:

\be \Delta \hat{S}(u) = 0  \ . \ee

\noindent A quick inspection of (\ref{StarTransform}) reveals that
only the even part of $\hat{S}(x)$,

\be \hat{S}_e(-x) = \hat{S}_e(x) \ , \ee

\noindent enters the formula for $\hat{S}(u)$. This means that for
$n$ even only $A_n$ and for $n$ odd only $B_n$ part of the
positive (negative) frequency modes enters the formula for
$\hat{S}(u)$. We obtain the following expression for $\hat{S}(u)$

\be \label{phaseLob} \hat{S}(u) = \hat{S}_0 +
{\sum_{nM}}(f_n(y)Y_{nM}(\textbf{n})\hat{a}_{nM} + h.c.) \ , \ee

\noindent where the sum is over $n\neq 0$. The $f_n(y)$ modes on
\L obaczewski space are the same functions of $y$ as before but
here we have $y = coth^2{\psi}$. The fact that only even modes in
$\hat{S}(x)$ appear in $\hat{S}(u)$ implies that $\hat{Q}$ does
not appear in it. The two-point correlation function of the vertex
operator

\be \hat{U}(u) = e^{i\hat{S}(u)} \ , \ee

\noindent is a finite function which does not require that the
normal ordering prescription be applied in marking contrast to the
de Sitter case

\be \label{correlLob} <0\mid e^{i\hat{S}(u)}e^{-i\hat{S}(v)}\mid
0> = e^{H(u,v)} \ , \ee

\be \label{Hfunct} H(u,v) = - \frac{1}{2}<0\mid (\hat{S}(u) -
\hat{S}(v))^2\mid 0> + \frac{1}{2}[\hat{S}(u),\hat{S}(v)] \ . \ee

\noindent One can show that the commutator function vanishes on \L
obaczewski space. Lorentz invariance implies that $H(u,v)$ is a
function of the scalar product $u{\cdot}v = cosh\lambda = z$,
$H(u,v) = H_d(z)$, which vanishes for $z = 1$ and satisfies an
inhomogeneous Laplace equation on \L obaczewski space:

\be \Delta_u H(u,v) = const  \ . \ee

\noindent The complete result of integration of the resulting
differential equation on $d$-dimensional \L obaczewski space is

\bea H_{d}(z) &=& - e^2
\frac{2^{d-3}(d-1)}{{\pi}(d-2)!}\Gamma^2(\frac{d-1}{2}) \times
 \,\nonumber\\&& \times{\int_1^z} dx (x^2 - 1)^{-\frac{d}{2}} {\int_1^x}
dy (y^2 - 1)^{\frac{d}{2} - 1} \ . \label{finresult}\eea

\noindent In particular for $d=3$ we find
\cite{as87,as89,as90,as92,as02,as04}

\be H_3(\lambda) = - {e^2\over \pi}({\lambda}coth{\lambda} - 1) \
, \ee

\noindent and for $d=4$ we have

\be H_4(\lambda) = - {e^2\over 2}(lncosh{\lambda\over 2} + {1\over
4 }tanh^2{\lambda\over 2}) \ . \ee

\noindent This completes our demonstration that the proper
framework for the discussion of the relation between correlation
functions on de Sitter (or \L obaczewski) space and correlation
functions of the Euclidean CFT on a time-like infinity $S^{d-1}$
are the generalized vertex operators and this is the context where
the so-called dS/CFT correspondence
\cite{as87,as89,pom90,pom91,amm97,mazmott,strom} was first
discovered as a simple side effect of work on the physical problem
\cite{as87,as89,as90,as92,as02,as04}.

\noindent \textbf{Acknowledgments}

The work of P. O. M. is partially supported by the NSF grant to
the University of South Carolina. One of us (A. C.) wishes to
acknowledge research support by U.S.C. during his sabbatical leave
of absence.

\end{document}

\end